\documentclass[amsfonts, amssymb, amsmath, reprint, twocolumn, superscriptaddress]{revtex4-1}

\usepackage{xcolor}
\usepackage{graphicx}
\usepackage{hyperref}
\usepackage{dcolumn}
\newcolumntype{.}{D{x}{}{13}}

\begin{document}


\title{Precision Spectroscopy of 
2S--$\boldsymbol{n}$S Transitions in Atomic Hydrogen: \linebreak 
A Determination of the Proton Charge Radius}

\author{R. G. Bullis}

\affiliation{Department of Physics, Colorado State University, Fort Collins,
Colorado, 80523, USA}

\author{W. L. Tavis}

\affiliation{Department of Physics, Colorado State University, Fort Collins,
Colorado, 80523, USA}

\author{M. R. Weiss}

\affiliation{Department of Physics, Colorado State University, Fort Collins,
Colorado, 80523, USA}

\author{J. Orellana Cisneros}

\affiliation{Department of Physics, Colorado State University, Fort Collins,
Colorado, 80523, USA}

\author{A. J. Cheeseman}

\affiliation{Department of Physics, Colorado State University, Fort Collins,
Colorado, 80523, USA}

\author{U. D. Jentschura}

\affiliation{Department of Physics and LAMOR, Missouri University of Science
and Technology, Rolla, Missouri, 65409, USA}

\author{D. C. Yost}

\affiliation{Department of Physics, Colorado State University, Fort Collins,
Colorado, 80523, USA}

\begin{abstract} 
We present absolute frequency measurements of
$2\text{S}_{1/2}-\textit{n}\text{S}_{1/2}$ two-photon transitions with
\textit{n}~=~8, 9, and 10 in a cryogenic beam of atomic hydrogen. Each
transition has been measured with a fractional uncertainty of $\approx$ 2.6
$\times~10^{-12}$. Combining the results from this work and the
$1\text{S}_{1/2}-2\text{S}_{1/2}$ transition frequency, we extract a
root-mean-square proton radius of $r_p$ = 0.8433(31) fm and a Rydberg frequency
of $cR_{\infty}~=~3\, 289\, 841\, 960\, 252.9 \, (9.7)~\text{kHz}$. These are
in good agreement with the CODATA 2022 recommended values.  
\end{abstract}

\maketitle

The hydrogen atom has long stood as a testbed of fundamental physics since its simple two-body nature allows for accurate theoretical calculations \cite{CODATA, KarshenboimQED, KarshenboimQED2}. These predictions can be compared with 
experiment \cite{Bullis, Parthey, Weitz, Nez, Biraben1, Biraben2, Diermaier, Beyer, Fleurbaey, Yzombard, Grinin, Brandt,Bezginov, Merkt2, Lothar} as tests of fundamental and Beyond Standard Model (BSM) physics \cite{Safronova, Karshenboim1, Karshenboim2, Jones, Potvliege, Frugiuele, Berengut}. Additionally, comparison of hydrogen and antihydrogen spectroscopy can stringently constrain charge-parity-time-reversal symmetry (CPT) violation \cite{Ahmadi1, Ahmadi2, Baker}. Of particular recent interest has been spectroscopic extractions of the proton charge radius since many of the results from hydrogen have shown disagreements from the very precise determinations from muonic hydrogen \cite{Pohl,Antognini, Pachucki}. While some recent measurements have shown good agreement with the muonic proton charge radius \cite{Bezginov, Beyer, Grinin, Merkt2}, there is tension with others \cite{Brandt, Fleurbaey}.

The $2\text{S}_{\text{1/2}}-n\text{S}_{\text{1/2}}$ two-photon transitions are attractive for precision measurement due to their relatively narrow natural linewidths ($\le$~144~kHz for n~$\ge$~8), resolved hyperfine structure, and greatly reduced sensitivity to Zeeman shifts. Here, we present absolute frequency measurements of the $2\text{S}_{1/2}-8\text{S}_{1/2}$, $2\text{S}_{1/2}-9\text{S}_{1/2}$, and $2\text{S}_{1/2}-10\text{S}_{1/2}$ transitions.  The uncertainty of the $2\text{S}_{1/2}-8\text{S}_{1/2}$ has been reduced by a factor of 4.4 \cite{Biraben1, Biraben2} while the two other transitions have never been previously measured.

\begin{figure}[h!]
\centering
\includegraphics[scale=0.4]{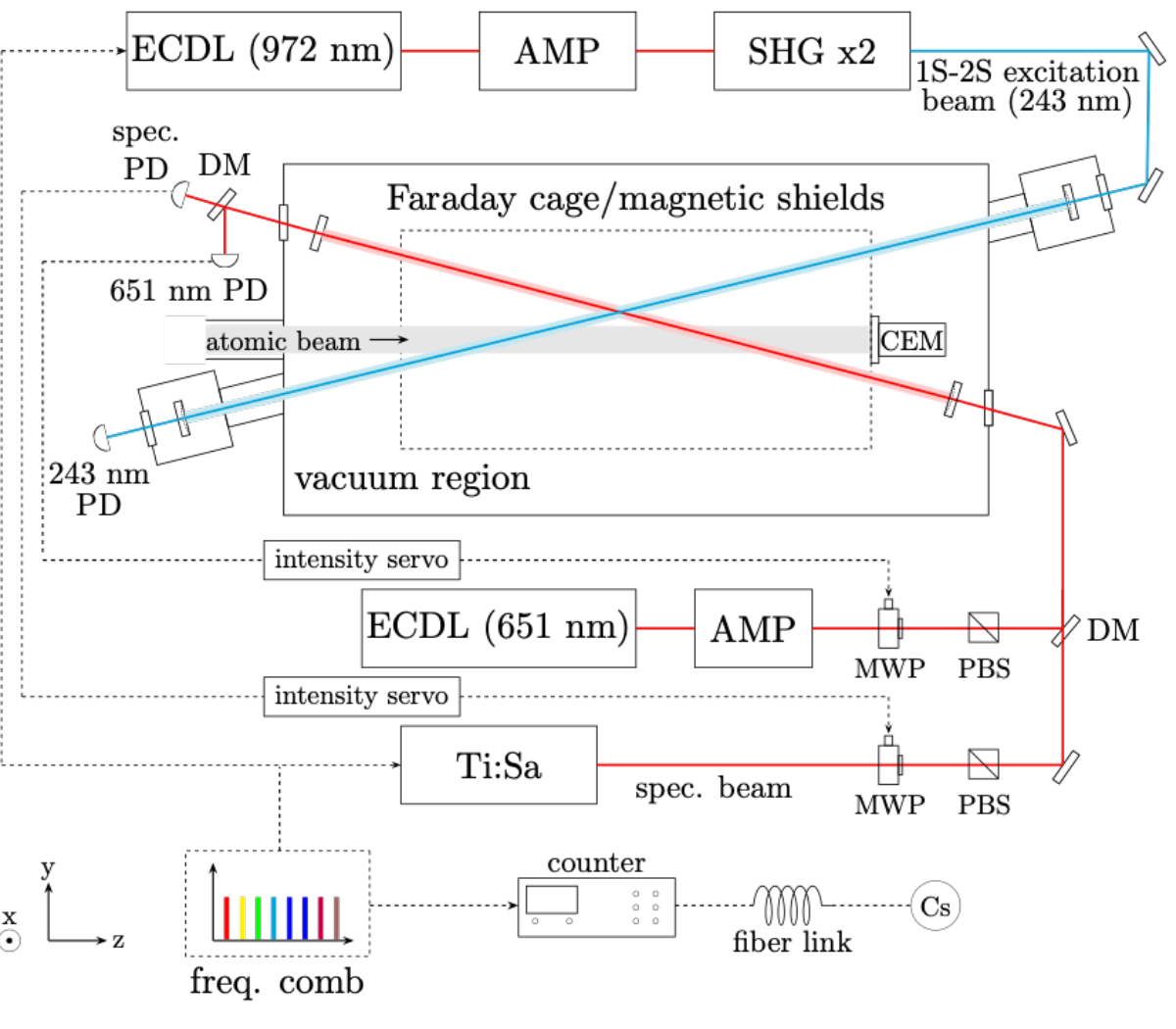}
    \caption{\label{fig1} Experimental apparatus. Photodiodes (PD) are used to monitor the intracavity power of both the spectroscopy and canceling lasers. These photodiodes are also used to provide the feedback signal for an intensity servo. Motorized waveplates (MWP) followed by polarizing beamsplitters (PBS) are used to control the input powers to the optical cavity. ECDL: external cavity diode laser. SHG: second harmonic generation. DM: dichroic mirror.}
\end{figure}

The experimental apparatus is shown in Fig.~\ref{fig1} and is similar to that in 
Ref.~\cite{Bullis2}. A highly collimated cryogenic beam of atomic hydrogen \cite{Scoop} is selectively excited to the metastable $2\text{S}_{1/2}(\text{F}=1)$ state using cavity enhanced 243 nm radiation \cite{Zak}. Once in the metastable state, the atoms travel $\approx$ 20 cm before interacting with cavity enhanced spectroscopy radiation (759-778 nm) and auxiliary radiation at 651.3 nm. The purpose of the radiation at 651.3 nm is to actively mitigate the effects of the ac Stark shift as described in Ref. \cite{Bullis2}. The spectroscopy laser cavity is at a $\approx$~6$^\circ$ angle with respect to the atomic beam to increase the interaction time. If the $2\text{S}_{1/2}(\text{F=1})-n\text{S}_{1/2}$(F=1) transition occurs, then atoms will decay to the ground state through a P state (the 2P is the most probable). The spectroscopic signal is the remaining metastable $2\text{S}_{1/2}$ atoms which are detected with a channel electron multiplier (CEM). The stabilization and frequency determination of the spectroscopy laser is performed in a manner similar to that in \cite{Brandt}.  However, we now have access to an ensemble of Cs beam clocks located at the NIST WWV radio station through a stabilized fiber link \cite{Vanarsdale} and the absolute frequency of our measurements is derived from these clocks \cite{supplement}.

We have performed a total of 48 blinded measurements (divided between the three transitions) over seven months. Each measurement run typically consists of $\approx$100 scans of the resonance. Individual scans are fit through a $\chi^2$ minimization procedure with a function of the form 
\begin{equation}
\label{lorentz}
F(\nu)= A -  \frac{B}{1+{4(\nu-\nu_c)^2/\Gamma^2}} \,,    
\end{equation}
where $A$ is an offset arising from the metastable count rate, and the second term is a Lorentzian function with amplitude $B$, linecenter $\nu_c$, and linewidth $\Gamma$. The negative sign before the Lorentzian arises since we are measuring depletion of the metastable signal.

\begin{figure}[h]
    \centering
    \includegraphics[scale=0.63]{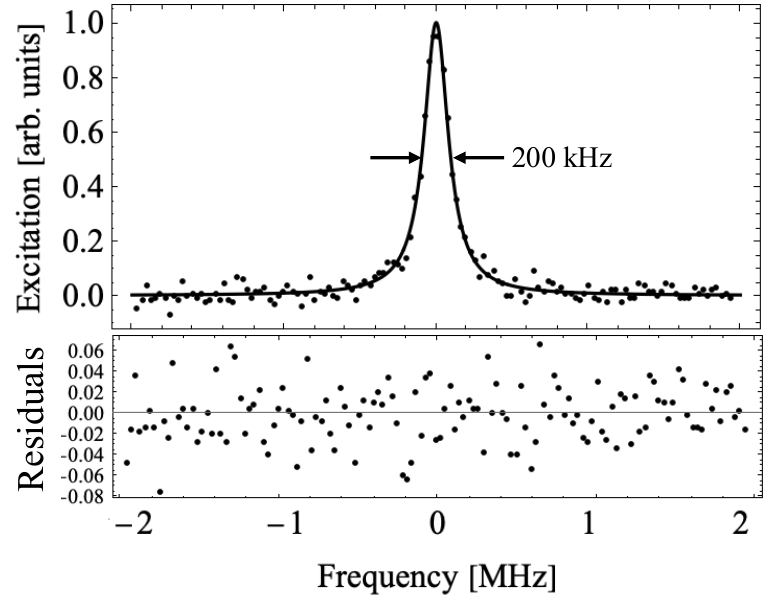}
    \caption{\label{fig2} \textit{Upper graph}: Experimentally obtained lineshape of the $2\text{S}_{1/2}-8\text{S}_{1/2}$ transition with active ac Stark shift mitigation fit with a Lorentzian function. \textit{Bottom graph}: The residuals of the fit. 
    }
\end{figure}

The most challenging systematic for 
these experiments is the characterization of the ac Stark shift, which typically has a magnitude of 100-400 kHz \cite{Brandt, Biraben1, Biraben2} 
and can distort the resonance~\cite{Bullis2}. 
These shifts are also nonlinear due to the Gaussian beam intensity profile and the presence of many atomic trajectories, all of which sample slightly different ac Stark shifts. Fortunately, with the introduction of active ac Stark shift mitigation 
demonstrated in \cite{Bullis2}, the total ac Stark shift is reduced by over an order of magnitude. In addition, 
the distortions of the atomic resonance are largely mitigated,
resulting in the recovery of Lorentzian lineshapes. A typical scan of the $2\text{S}_{\text{1/2}}-8\text{S}_{\text{1/2}}$ transition fit with 
Eq.~\eqref{lorentz} is shown in Fig.~\ref{fig2}.

To guide our data analysis, we have developed a lineshape model, which takes into account the experimental geometry, the atom-laser excitation dynamics from both the spectroscopy and ac Stark shift canceling lasers, and the metastable velocity distribution. Our model indicates that the recovered lines should be well-described by Eq.~\eqref{lorentz} and our experimental results confirm this (see Fig.~\ref{fig2}).
This model is also used to simulate the ac Stark shift extrapolations with both the spectroscopy and ac Stark shift mitigating radiation. We find that the ac Stark shift is well-described by a linear function of both laser powers such that 
\begin{equation}
\label{model1}
\nu_c = \nu_0+\beta_1 P_{\beta_1}-\beta_2 P_{\beta_2},
\end{equation}
where $P_{\beta_1}$ and $P_{\beta_2}$ are the spectroscopy and ac Stark shift mitigating laser powers respectively, $\beta_1$ and $\beta_2$ are positive fit parameters, and $\nu_0$ is the linecenter at zero laser power. A small nonlinearity remains which can produce offsets of up to 100 Hz in our extrapolations which is accounted for in our error budget~\cite{supplement}. However, without ac Stark shift mitigation, the nonlinearity is more than an order-of-magnitude larger  \cite{Brandt} which highlights an additional major advantage of this method.

We have implemented an intensity stabilization servo to maintain the ratio between the spectroscopy and canceling laser powers. In this case, the two dimensional ac Stark shift extrapolation becomes an effective one dimensional extrapolation where
\begin{equation}
\label{model2}
    \nu_c = \nu_0 + \gamma P_{\beta_1}, 
\end{equation}
with $\gamma = \beta_1-\beta_2\, P_{\beta_2}/P_{\beta_1}$. Typically, the power ratio is chosen such that $|\gamma|< 0.1 \times \beta_1$. With the intensity stabilization servo, the power ratio can be stably maintained to within a few percent over the course of data collection. While Eq.~\eqref{model2} allows for easier data visualization, in practice, data is fit with Eq.~\eqref{model1} to ensure no errors are introduced through imperfections in the intensity stabilization system. However, we have found that the obtained zero power linecenter, $\nu_0$, is insensitive to which data analysis method is used. 

\begin{figure}[h]
    \centering
    \includegraphics[scale=0.88]{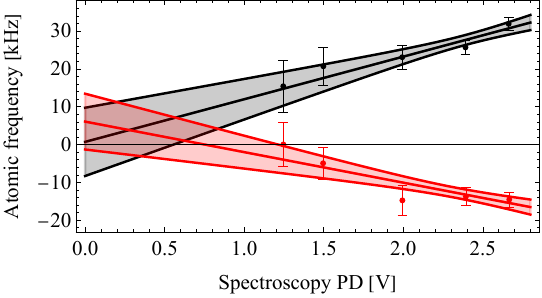}
    \caption{\label{fig3} ac Stark shift extrapolations for the $2\text{S}_{\text{1/2}}-8\text{S}_{\text{1/2}}$ transition with an undercompensated (black, $\chi_r^2 = 1.05$) and overcompensated ac Stark shift (red, $\chi_r^2 = 1.01$). Extrapolations are performed using Eq. 3. Each individual point consists of approximately 10 scans of the resonance. Frequency values are listed relative to the theoretical value calculated using the CODATA 2022 recommended fundamental constants.}
\end{figure}

From our numerical model, we find the extrapolations are sufficiently linear whether the ac Stark shift is slightly over or undercompensated.  However, the linearity is slightly better with an undercompensated ac Stark shift (positive $\gamma$), so extrapolations are typically performed under those conditions~\cite{supplement}.
As an experimental check, we have performed extrapolations with both signs of $\gamma$ to ensure that results are consistent. Example extrapolations are shown in 
Fig.~\ref{fig3} for the $2\text{S}_{\text{1/2}}-8\text{S}_{\text{1/2}}$ transition. 
For each day of data collection, a new extrapolation is performed with no prior knowledge of $\beta_1$ or $\beta_2$ assumed.

The second leading systematic is the dc Stark effect. Stray electric fields are passively mitigated by coating all surfaces surrounding the spectroscopy region with colloidal graphite \cite{Porter}. In several similar experiments, dc Stark shifts are characterized by measuring high lying $2\text{S}_{1/2}-n\text{D}_{5/2}$ transitions with $n \ge 12$ (see Refs.~\cite{Brandt, Biraben1, Biraben2}). These transitions distort due to the quasidegenerate $n\text{D}_{5/2}$ and $n\text{F}_{5/2}$ states, and this distortion is used to characterize both the direction and magnitude of the electric field. Using this method, we have found electric fields to be of magnitude 0.418(30)~V/m \cite{supplement} and are constant in time as long as the apparatus remains under vacuum \cite{Brandt}.

We also characterize stray fields through absolute frequency measurements of the $2\text{S}_{\text{1/2}}-16\text{S}_{\text{1/2}}$ transition. The sensitivity of the $n\text{S}_{1/2}$ states to stray electric fields scale as $\approx n^7$ so that the $2\text{S}_{\text{1/2}}-16\text{S}_{\text{1/2}}$ transition is $\approx$~128 times more sensitive than the $2\text{S}_{\text{1/2}}-8\text{S}_{\text{1/2}}$ transition. 
However, transitions involving the 2S$_{1/2}$ are 
sensitive to nuclear structure effects ({\em i.e.}, the proton charge radius). 
Therefore, we extract the dc electric field from the shift of the $10\text{S}_{1/2}-16\text{S}_{1/2}$ interval, 
which is $\approx$ 125 less sensitive to the proton finite-size effect 
than transitions involving the $2\text{S}_{1/2}$ state.
The $10\text{S}_{1/2}-16\text{S}_{1/2}$
shifts at a rate of 967.388~$\text{kHz} / \text{(V/m)}^2$
in a static electric field, and its dc Stark shift greatly exceeds the residual
proton-size effect on the $10\text{S}_{1/2}$ state at typical electric field strengths.
Combining our measurements of the $2\text{S}_{\text{1/2}}-16\text{S}_{\text{1/2}}$ and $2\text{S}_{\text{1/2}}-10\text{S}_{\text{1/2}}$ transitions, we find a shift of the $10\text{S}_{\text{1/2}}-16\text{S}_{\text{1/2}}$ interval from the theory by 166(22)~kHz and, from this, determine the magnitude of the electric field to be 0.415(27)~V/m, which is in agreement with the determinations from $2\text{S}_{1/2}-n\text{D}_{5/2}$ distortion data \cite{supplement}. 

\begin{figure}[h]
    \centering
    \includegraphics[scale=0.41]{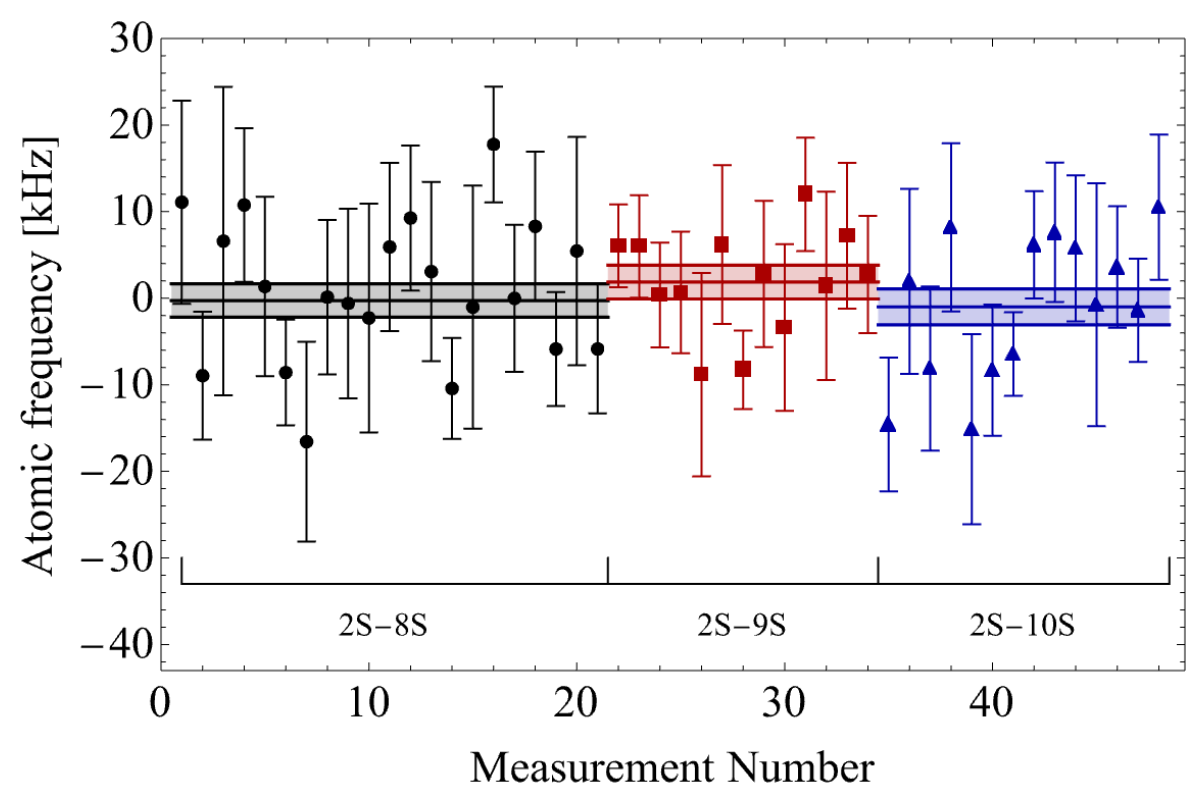}
    \caption{\label{fig4} dc Stark shift corrected extrapolations of the $2\text{S}_{\text{1/2}}-8\text{S}_{\text{1/2}}$ ($\chi_r^2 = 1.06$), $2\text{S}_{\text{1/2}}-9\text{S}_{\text{1/2}}$ ($\chi_r^2 = 0.94$), and $2\text{S}_{\text{1/2}}-10\text{S}_{\text{1/2}}$ ($\chi_r^2 = 1.04$) hyperfine centroid transition frequencies. The values are relative to the theoretical transition frequency as calculated using the CODATA 2022 recommended fundamental constants.}
\end{figure}

Each ac Stark shift extrapolated value is corrected for the dc Stark shift using the theoretically calculated dc Stark shift coefficients of 7.768 $\text{kHz} / \text{(V/m)}^2$, 17.691 $\text{kHz} / \text{(V/m)}^2$, and 37.242 $\text{kHz} / \text{(V/m)}^2$ for the $2\text{S}_{\text{1/2}}-8\text{S}_{\text{1/2}}$, $2\text{S}_{\text{1/2}}-9\text{S}_{\text{1/2}}$, and $2\text{S}_{\text{1/2}}-10\text{S}_{\text{1/2}}$ transitions, respectively. To test the consistency of this procedure, the entire system was recoated in colloidal graphite and the fields re-characterized using the $10\text{S}_{1/2}-16\text{S}_{1/2}$ interval. In this case, we found the deviation from theory to be 115(8.7)~kHz, giving a residual electric field magnitude of 0.345(13)\,${\rm V}/{\rm m}$. Spectroscopy data  taken before and after recoating was consistent for all three transitions measured. All dc Stark shift corrected extrapolations are shown in Fig.~\ref{fig4}. The weighted dc electric field corrections and errors are given in Table~I.

\setlength{\tabcolsep}{5pt} 

\begin{table*}[ht]
\begin{center}
\begin{minipage}{0.9\linewidth}
\begin{center}
\caption{\label{table1} Systematic corrections to the $2\text{S}_{1/2}$--$n\text{S}_{1/2}$ hyperfine 
centroid frequencies in atomic hydrogen. All values are given in kHz.}
\centering
\def\arraystretch{1.15}
\begin{tabular}{l...}
\hline
\hline
& \multicolumn{1}{c}{$\nu(2\text{S}_{1/2}-8\text{S}_{1/2})$}
& \multicolumn{1}{c}{$\nu(2\text{S}_{1/2}-9\text{S}_{1/2})$}
& \multicolumn{1}{c}{$\nu(2\text{S}_{1/2}-10\text{S}_{1/2})$} \\
\hline
ac Stark shift corrected 
  & \multicolumn{1}{r}{770\,649\,350\,002.60(1.90)}
  & \multicolumn{1}{r}{781\,432\,010\,264.30(1.89)}
  & \multicolumn{1}{r}{789\,144\,778\,098.87(2.05)} \\
\hline
dc Stark shift             & -1.x30 (0.17) & -2.x90 (0.40) & -5.x89(0.83) \\
Second-order Doppler shift &  0.x74 (0.10) &  0.x75 (0.10) &  0.x76(0.10) \\
Blackbody radiation        & -0.x39 (0.02) & -0.x76 (0.03) & -1.x07(0.04) \\
Extrapolation nonlinearity & 0.x00(0.10)   &  0.x00(0.10)  &  0.x00(0.10) \\
Photodiode imperfections   & 0.x00(0.28)   &  0.x00(0.28)  &  0.x00(0.28) \\
Frequency calibration      & 0.x00(0.08)   &  0.x00(0.08)  &  0.x00(0.08) \\
ac Stark canceling laser stability
                           & 0.x00(0.03)   & 0.x00(0.03)   & 0.x00(0.03)\\
\hline
Final measured value
   & \multicolumn{1}{r}{770\,649\,350\,001.65(1.94)} 
   & \multicolumn{1}{r}{781\,432\,010\,261.39(1.96)} 
   & \multicolumn{1}{r}{789\,144\,778\,092.67(2.24)} \\
\hline
\hline
\end{tabular}
\end{center}
\end{minipage}
\end{center}
\end{table*}%

To determine the 2\textsuperscript{nd}-order Doppler shift, we have measured the velocity distribution of our atomic beam using an in-vacuum atomic beam chopper and recording the time-of-flight distribution (similar to \cite{Scoop}). From the time-of-flight data, we obtain a velocity distribution which is used to calculate an average shift and uncertainty. The results vary minimally if a shift corresponding only to the most probable velocity (410~m/s) is used. The blackbody radiation shifts have been theoretically calculated in \cite{Farley} and the correction is assigned for a temperature of 300(5)~K. Many of these shifts have recently been recalculated with fine structure, and have shown good agreement \cite{Pot}. 

A summary of hyperfine centroid frequencies with systematic corrections for all three transitions is given in Table 1. Pressure shifts and Zeeman shifts are estimated to be very small in our apparatus ($\leq1$ Hz) and so are omitted from the table. Further information regarding these and other systematic corrections are given in \cite{supplement}. Our measurement of the $2\text{S}_{1/2}-8\text{S}_{1/2}$ transition is in a combined $1.2~\sigma $ agreement with the previous measurement of this transition 
and improves upon the uncertainty by a factor of 4.4 \cite{Biraben2}. Our spectroscopy of the $2\text{S}_{1/2}-9\text{S}_{1/2}$ and $2\text{S}_{1/2}-10\text{S}_{1/2}$ transitions represent the first precision measurements of these transitions to date. 

\begin{figure}[!h]
    \centering
    \includegraphics[scale=0.62]{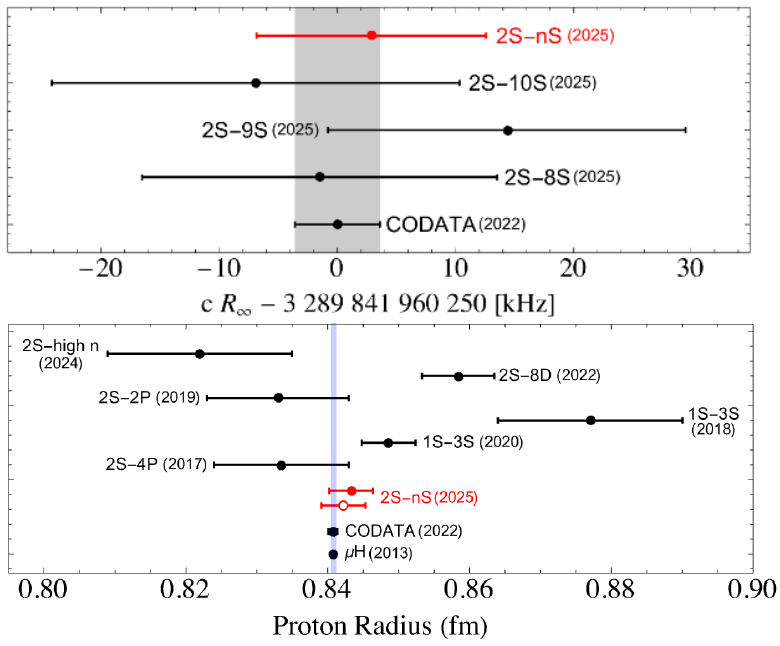}
    \caption{\label{fig5}\textit{Upper graph:} Rydberg frequencies with 1$\sigma$ uncertainties extracted from the measurements of the $2\text{S}_{1/2}-8\text{S}_{1/2}$, $2\text{S}_{1/2}-9\text{S}_{1/2}$, $2\text{S}_{1/2}-10\text{S}_{1/2}$ transitions relative to the CODATA 2022 recommended value. 
    \textit{Bottom graph:} A selection of proton radius extractions from hydrogen spectroscopy since 2010. 2S-4P: \cite{Beyer}, 1S-3S (2018): \cite{Fleurbaey}, 2S-2P: \cite{Bezginov}, 1S-3S (2020): \cite{Grinin}, 2S-8D: \cite{Brandt}, 2S-high $n$: \cite{Merkt2}.  Error bars represent the 1$\sigma$ uncertainty for each measurement and for transitions where $\Delta n \neq 0$, the 1S-2S transition is used in the extraction \cite{Parthey}. 
    The proton radius extracted from muonic hydrogen spectroscopy is given by the light blue bar. The proton radius from this work is given in red. The red filled circle corresponds to the result given in 
    Eq.~\eqref{rp}, whereas the open circle uses the status of the theory in Ref. \cite{CODATA}.}
\end{figure}

Of particular interest is the use of these transition frequencies to extract a proton charge radius and a Rydberg constant. For this, we require the
$\text{1S}_{1/2}-2\text{S}_{1/2}$ transition frequency from \cite{Parthey}
along with the bound-state QED corrections summarized in \cite{CODATA}. However, several corrections for $n$S states with $n>8$ have not been
previously calculated. Therefore, we present relativistic Bethe logarithms, $A_{60}(n\text{S})$, for $n=9,10$
in Table~II (for the definition of $A_{60}$, 
see Eq.~(3.43) and Table~1 of Ref.~\cite{jentschura}). Furthermore, we have
obtained results for the self-energy remainder function $G_{\rm SE}(Z \alpha)$
at $Z=1$, and perform a reevaluation of  the vacuum-polarization remainder
function $G_{\rm VP}(Z\alpha)$ (for the notation, see Table~VII of
Ref.~\cite{CODATA}). The results obtained for the latter are
consistent with recent analytic investigations on higher-order
vacuum-polarization corrections discussed in Ref.~\cite{Karr}. All of these
results are presented in Table~II.  
Taking these into account, together with
the recently improved two-loop self-energy corrections for the hydrogen ground
state~\cite{Yerokhin}, we extract a Rydberg
constant and proton radius for each transition measured. By taking a weighted average and accounting for correlated errors 
between our measurements \cite{Lista}, we obtain
\begin{align}
\label{rp}
    r_p =& \; 0.8433(31) \text{ fm},  \\ 
    cR_{\infty} =& \; 3~289~841~960~252.9(9.7)~\text{kHz}.
\end{align}
These results and those of several other recent determinations of the proton radius are summarized in Fig.~\ref{fig5}. 

The recent theoretical advances in two-loop corrections reported in Ref.~\cite{Yerokhin} were not available for the earlier measurements shown in Fig.~\ref{fig5}. Therefore, to provide a consistent comparison between experimental results, we also calculate a proton radius using the status of theory as given in Ref.~\cite{CODATA}. This gives  $r_p = 0.8422(31)~\text{fm}$.  This value is also included in Fig.~5 marked with a red open circle.  Our extraction of the Rydberg constant is not significantly altered by the theory presented in \cite{Yerokhin}.

\begin{table}[ht]
\begin{center}
\begin{center}
\caption{\label{table2} Calculations of relativistic Bethe logarithm $A_{60}$, self-energy remainder function $G_{\text{SE}}$, and vacuum-polarization remainder function $G_{\text{VP}}$. All values are unitless.}
\centering
\def\arraystretch{1.15}
\begin{tabular}{lc c c}
\hline
\hline
& \multicolumn{1}{c}{$A_{60}$}
& \multicolumn{1}{c}{$G_{\text{SE}}$}
& \multicolumn{1}{c}{$G_{\text{VP}}$} \\
\hline
$8\text{S}_{1/2}$ & $-31.264257(1)$ &$-30.620(7)$ & $-0.78120(1)$ \\
 $9\text{S}_{1/2}$ & $-31.224646(1)$   &  $-30.582(7)$ & $-0.77738(1)$  \\
    $10\text{S}_{1/2}$ & $-31.191983(1)$  &  $-30.550(7)$ & $ -0.77418(1)$ \\
\hline
\hline
\end{tabular}
\end{center}
\end{center}
\end{table}%

In conclusion, we present measurements of the $2\text{S}_{1/2}-8\text{S}_{1/2}$, $2\text{S}_{1/2}-9\text{S}_{1/2}$, and $2\text{S}_{1/2}-10\text{S}_{1/2}$ transitions, each with a fractional uncertainty of $\approx$~2.6~$\times~10^{-12}$. These measurements can be combined with the measured $1\text{S}_{1/2}-2\text{S}_{1/2}$ transition to give a proton charge radius and Rydberg constant which is in good agreement with the CODATA 2022 recommended value. 
The proton charge radius and Rydberg constant from this work disagrees with that extracted from our measurement of the $2\text{S}_{1/2}-8\text{D}_{5/2}$ transition by $\approx 2.5~\sigma$ \cite{Brandt}. 
During the course of the measurements presented here, we have not been able to find a likely theoretical issue or underestimated systematic effects which would explain this discrepancy. However, we note that possible systematic effects for $2\text{S}_{1/2}-n\text{S}_{1/2}$ measurements are reduced as compared with the $2\text{S}_{1/2}-n\text{D}_{5/2}$. In particular, for a $2\text{S}_{1/2}-n\text{D}_{\text{5/2}}$ measurement, hyperfine structure of the $n\text{D}_{\text{5/2}}$ level is unresolved and the excitation ratio between the hyperfine levels must be known. Further, the quasidegeneracy of the $n\text{D}_{\text{5/2}}$ and $n\text{F}_{\text{5/2}}$ states produces a linear Stark shift which can distort the resonant lineshape. Finally, Zeeman shifts are polarization dependent and amount to a $\approx$~3~kHz/mGauss shift with circular polarization. This can be compared with $2\text{S}_{1/2}-n\text{S}_{1/2}$ measurements presented here which have well separated hyperfine structure, no near degenerate states of opposite parity, and negligible Zeeman shifts \cite{supplement}.

Our results are especially pertinent to BSM theories which introduce light bosons \cite{Karshenboim1, Karshenboim2, Jones, Frugiuele}.  These bosons modify the Coloumb potential resulting in variations of the Rydberg constant based on the principle quantum numbers of the transitions used in the extractions.  Since we measure three transitions in the same apparatus, with high precision and varying $n$, our results could be used to constrain such theories. There is also a strong motivation for extending our $2\text{S}_{1/2}-n\text{S}_{1/2}$ measurements to deuterium since the isotope shifts for these transitions can provide a sensitive test of BSM physics \cite{Potvliege}. 

{\em Note added.---}Shortly before acceptance of this manuscript, 
a high precision measurement of the hydrogen
2S-6P transition was published~\cite{AddedInProof}.

{\em Acknowledgements.---}We would like to thank 
Thomas Udem, Randolf Pohl, Vitaly Wirthl, Lothar
Maisenbacher, Omer Amit, Derya Taray, Vincent Weis,  Jacob Roberts, Sam Brewer,
and Christian Sanner for insightful discussions. We would also like to thank
Glenn Nelson, Matt Deutch, Bill Yates, Jim Spicer, Michael Lombardi, Judah
Levine, and Jeffrey Sherman at NIST for assistance with the system of cesium
beam clocks used for frequency calibration. This research has been supported by
the National Science Foundation (PHY-2207298 and PHY-2513220).

{\it Data availability.---}The data that support the findings of this article
are openly available \cite{DataAvailability}.


\begin{thebibliography}{}

\bibitem{CODATA} P. J. Mohr, D. B. Newell, B. N. Taylor, and E. Tiesinga,
``CODATA recommended values of the fundamental physical constants: 2022'', Rev.
Mod. Phys. {\bf 97}, 025002 (2025).

\bibitem{KarshenboimQED}
S. G. Karshenboim, ``Precision physics of simple atoms: QED tests, nuclear structure and fundamental constants'', Phys. Rep. 422, \textbf{1} (2005).

\bibitem{KarshenboimQED2} S. G. Karshenboim, F. S. Pavone, G. F. Bassani, M.
Inguscio, T. W. H\"{a}nsch, ``Introduction to Simple Atoms''
(Springer, New York, 2007).

\bibitem{Yzombard} P. Yzombard, \textit{et. al.}, ``1S--3S cw spectroscopy of
hydrogen/deuterium atom'' Eur. Phys. J. \textbf{77} 2 (2023).

\bibitem{Parthey} C. G. Parthey, A. Matveev, J. Alnis, B. Bernhardt, A. Beyer,
R. Holzwarth \textit{et al.}, ``Improved Measurement of the Hydrogen 1S--2S
Transition Frequency,'' Phys. Rev. Lett. \textbf{107}, 203001 (2011).

\bibitem{Nez} F. Nez, \textit{et al.}, ``Precise frequency measurement of the
$2\text{S}_{\text{1/2}}$--$8\text{S}_{\text{1/2}}$/8D
transtions in atomic hydrogen: New determination of
the Rydberg constant,'' Phys. Rev. Lett.  \textbf{69}, 2326 (1992).

\bibitem{Weitz} M. Weitz, \textit{et al.}, ``Precision measurement of the 1S
ground-state Lamb shift in atomic hydrogen and deuterium by frequency
comparison'' Phys. Rev. A \textbf{52}, 2664 (1995).

\bibitem{Biraben1} B. de Beauvoir, F. Nez, L. Julien, B. Cagnac, F. Biraben, D.
Touahri, L. Hilico, O. Acef, A. Clairon, and J. J. Zondy, ``Absolute frequency
measurement of the $2\text{S}$--$8\text{S}$/8D 
transitions in hydrogen and deuterium: new determination of the Rydberg
constant'', Phys. Rev. Lett., \textbf{78}, 440 (1997). 

\bibitem{Biraben2} B. de Beauvoir, C. Schwob, O. Acef, L. Jozefowski, L.
Hilico, F. Nez, L.  Julien, A. Clairon, and F. Biraben, ``Metrology of the
hydrogen and deuterium atoms: Determination of the Rydberg constant and lamb
shifts'', Eur. Phys. J. D \textbf{12}, 61 (2000).

 \bibitem{Diermaier} M. Diermaier, C. B. Jepsen, B. Kolbinger, C. Malbrunot, O.
Massiczek, C. Sauerzopf, M. C. Simon, J. Zmeskal, and E. Widmann, ``In-beam
measurement of the hydrogen hyperfine splitting and prospects for antihydrogen
spectroscopy'', Nat. Comm. \textbf{8}, 15749 (2017).

\bibitem{Beyer} 
 A. Beyer, \textit{et al.}, ``The Rydberg constant and proton size from atomic hydrogen,'' Science \textbf{358}, 79 (2017).
 
\bibitem{Fleurbaey} H. Fleurbaey, S. Galtier, S. Thomas, M. Bonnaud, L. Julien,
F. Biraben, F. Nez, M. Abgrall, and J. Guena,``New Measurement of the 1S-3S
Transition Frequency of Hydrogen: Contribution to the Proton Charge Radius
Puzzle,'' Phys. Rev. Lett. \textbf{120}, 183001 (2018).

 \bibitem{Bezginov} N. Bezginov, T. Valdez, M. Horbatsch, A. Marsman, A. C.
Vutha, and E. A. Hessels, ``A measurement of the atomic hydrogen Lamb shift and
the proton charge radius'', Science \textbf{365}, 1007 (2019).

\bibitem{Grinin} A. Grinin, A. Matveev, D. C. Yost, L. Maisenbacher, V. Wirthl,
R. Pohl, T. W. Hänsch, and T. Udem, ``Two-photon frequency comb spectroscopy of
atomic hydrogen'', Science \textbf{370}, 1061 (2020).

\bibitem{Brandt} A. D. Brandt, S. F. Cooper, C. Rasor, Z. Burkley, A. Matveev,
and D. C. Yost, ``Measurement of the $2S_{1/2}-8D_{5/2}$ transition in
hydrogen'', Phys. Rev. Lett. \textbf{128}, 023001 (2022).

\bibitem{Merkt2} S. Scheidegger and F. Merkt, ``Precision-spectroscopic
determination of the binding energy of a two-body quantum system: the hydrogen
atom and the proton-size puzzle'', Phys. Rev. Lett. \textbf{132} 113001 (2024).

\bibitem{Lothar} L. Maisenbacher, ``Precision spectroscopy of the 2S--nP
transitions in atomic hydrogen'', Doctoral Thesis, University of Munich,
2021.

\bibitem{Bullis} R. G. Bullis, C. Rasor, W. L. Tavis, S. A. Johnson, M. R.
Weiss, and D. C. Yost, ``Ramsey spectroscopy of the 2S hyperfine interval in
atomic hydrogen'' Phys. Rev. Lett. \textbf{130}, 203001 (2023).

\bibitem{Safronova} M. S. Safronova, D. Budker, D. DeMille, D. F. Jackson
Kimball, A. Derevianko, and C. W. Clark, ``Search for new physics with atoms
and molecules'', Rev. Mod. Phys. \textbf{90}, 025008 (2018).

\bibitem{Karshenboim1} S. G. Karshenboim, ``Precision physics of simple atoms
and constraints on a light boson with ultraweak coupling'', Phys. Rev. Lett.
\textbf{104}, 220406 (2010).

\bibitem{Karshenboim2} S. G. Karshenboim, ``Constraints on a long-range
spin-independent interaction from precision atomic physics'', Phys. Rev. D
\textbf{82}, 073003 (2010).

\bibitem{Jones} M. P. A. Jones, R. M. Potvliege, and M. Spannowsky, ``Probing
new physics using Rydberg states of atomic hydrogen'', Phys. Rev. Research
\textbf{2}, 013244 (2020).

\bibitem{Frugiuele} C. Frugiuele, E. Fuchs, G. Perez, and M. Schlaffer,
``Constraining new physics models with isotope shift spectroscopy'',
Phys. Rev. D \textbf{96}, 015011 (2017).

\bibitem{Berengut} J. C. Berengut, D. Budker, C. Delaunay, V. V. Flambaum, C.
Frugiuele, E. Fuchs, C. Grojean, R. Harnik, R. Ozeri, G. Perez, and Y. Soreq,
``Probing new long-range interactions by isotope shift spectroscopy'', Phys.
Rev. Lett. \textbf{120}, 091801 (2018).

\bibitem{Potvliege} R. M. Potvliege, A. Nicolson, M. P. A. Jones, and M.
Spannowsky, ``Deuterium spectroscopy for enhanced bounds on physics beyond the
standard model'', Phys. Rev. A \textbf{108}, 052825 (2023).

\bibitem{Ahmadi1} M. Ahmadi \textit{et al.}, ``Observation of the 1S–2S
transition in trapped antihydrogen'', Nature (London) \textbf{541}, 506 (2017).

\bibitem{Ahmadi2} M. Ahmadi \textit{et al.}, ``Characterization of the 1S–2S
transition in antihydrogen'', Nature (London) \textbf{557}, 71 (2018).

\bibitem{Baker} C. J. Baker, W. Bertsche, A. Capra.  \textit{et al.},
``Precision spectroscopy of the hyperfine components of the 1S–2S transition in
antihydrogen,'' Nat. Phys. \textbf{21}, 201 (2025).

\bibitem{Pohl} R. Pohl \textit{et al.}, ``The size of the proton,'' Nature
(London) \textbf{466}, 213 (2010).

\bibitem{Antognini} A. Antognini \textit{et al.}, ``Proton structure from the
measurement of 2S-2P transition frequencies of muonic hydrogen,'' Science
\textbf{339}, 417 (2013).

 \bibitem{Pachucki} K. Pachucki \textit{et. al.,} ``Comprehensive theory of the
Lamb shift in light muonic atoms,'' Rev. Mod Phys, \textbf{96} 015001 (2024).

\bibitem{Bullis2} R. G. Bullis, W. L. Tavis, M. R. Weiss, D. C. Yost, ``Narrow
resonances in Rydberg hydrogen spectroscopy'', Phys. Rev. A. \textbf{110},
052807 (2024).

\bibitem{Scoop} S. F. Cooper, A. D. Brandt, C. Rasor, Z. Burkley, and D. C.
Yost, ``Cryogenic atomic hydrogen beam apparatus with velocity
characterization'', Rev. Sci. Instrum. \textbf{91}, 013201 (2020)

\bibitem{Zak} Z. Burkley, A. D. Brandt, C. Rasor, S. F. Cooper, and D. C. Yost,
``Highly coherent, watt-level deep-UV radiation via a frequency-quadrupled
Yb-fiber laser system,'' Appl. Opt. \textbf{58}, 1657 (2019).

\bibitem{Vanarsdale} J. B. VanArsdale, M. J. Deutch, M. A. Lombardi, G. K.
Nelson, J. A. Sherman, J. Spicer, W. C. Yates, D. C. Yost, and S. M. Brewer,
``Dissemination of UTC(NIST) over 20 km of commercial optical fiber with active
phase stabilization,'' Opt. Lett. \textbf{49}, 2545 (2024).

\bibitem{supplement} See Supplemental Material 
at \url{http://link.aps.org/supplemental/10.1103/lgl2-6cb8} 
for additional information
regarding systematic characterization, which includes Refs. \cite{Adhikari,
Matveev1, Matveev2, Haas}.

\bibitem{Adhikari} C. M. Adhikari and U. D. Jentschura, ``Long-range
interactions for hydrogen atoms in excited D states'', Atoms [MDPI]
\textbf{10}, 6 (2022).

\bibitem{Matveev1} U. D. Jentschura, C. M. Adhikari, R. Dawes, A. Matveev, and
N. Kolachevsky, ``Pressure shifts in high-precision hydrogen spectroscopy: I.
Long-range atom–atom and atom–molecule interactions'', J. Phys. B \textbf{52},
075005 (2019).

\bibitem{Matveev2} A. Matveev, N. Kolachevsky, C. M. Adhikari, U. D.
Jentschura, ``Pressure shifts in high-precision hydrogen spectroscopy: II.
Impact approximation and Monte-Carlo simulations'',  J. Phys. B \textbf{52},
075006 (2019).

\bibitem{Haas} M. Haas \textit{et al.}, ``Two-photon excitation dynamics in
bound two-body Coulomb systems including ac Stark shift and ionization'', Phys.
Rev. A \textbf{73}, 052501 (2006).

\bibitem{Porter} B. H. Porter, ``Research applications of colloidal graphite'',
Rev. Sci. Instrum. \textbf{7}, 101 (1936).

\bibitem{Farley}
J. W. Farley and W. H. Wing, ``Accurate calculation of dynamic Stark shifts and depopulation rates of Rydberg energy levels induced by blackbody radiation. Hydrogen, helium, and alkali-metal atoms,'' Phys. Rev. A, \textbf{23}, 2397 (1981).

\bibitem{Pot} R. M. Potvliege, private communication (2025).

\bibitem{Lista} L. Lista, "Combination of measurements and the BLUE method" EPJ
Web of Conferences \textbf{137}, 11006 (2017).

\bibitem{Yerokhin} V. A. Yerokhin, Z. Harman, and C. H. Keitel, ``Two-Loop
Electron Self-Energy for Low Nuclear Charges'', Phys. Rev. Lett. \textbf{133},
251803 (2024).

\bibitem{jentschura} U. D. Jentschura, A. Czarnecki and K. Pachucki,
``Nonrelativistic QED approach to the Lamb shift'', Phys. Rev. A \textbf{72},
062102 (2005).

\bibitem{Karr} J.-Ph. Karr, L. Hilico, and V. I. Korobov, ``One-loop vacuum
polarization at $m \alpha^7$ order for the two-center problem'', Phys. Rev. A
{\bf 90}, 062516 (2014).

\bibitem{AddedInProof} L. Maisenbacher, V. Wirthl, A. Matveev, A. Grinin, R.
Pohl, T. W. H\"{a}nsch, and Th. Udem, ``Sub-part-per-trillion test of the
standard model with atomic hydrogen'', Nature (London) \textbf{650}, 845
(2026).

\bibitem{DataAvailability} Supporting data for ``Precision spectroscopy of
2S-nS transitions in atomic hydrogen: a determination of the proton charge
radius'' is available at \url{http://doi.org/10.5281/zenodo.18747148}.

\end{thebibliography}
\end{document}